# Graphene based widely-tunable and singly-polarized pulse generation with random fiber lasers

B. C. Yao[1,2], Y. J. Rao[1*], Z. N. Wang[1], Y. Wu[1], J. H. Zhou[3], H. Wu[1], M. Q. Fan[1], X. L. Cao[1], W. L. Zhang[1], Y. F. Chen[3], Y. R. Li[3], D. Churkin[4,5,6], S. Turitsyn[4,5], and C. W. Wong[2]

[1]*Key Laboratory of Optical Fiber Sensing and Communications (Education Ministry of China), University of Electronic Science and Technology of China, Chengdu 610054, China*

[2] *Mesoscopic Optics and Quantum Electronics Laboratory, University of California, Los Angeles, CA 90095, United States*

[3]*State Key Laboratory of Electronic Thin Films and Integrated Devices, University of Electronic Science and Technology of China, Chengdu 610054, China*

[4]*Aston Institute of Photonic Technologies, Aston University, Birmingham, B47ET, United Kingdom*

[5]*Laboratory of Nonlinear Photonics, Novosibirsk State University, Novosibirsk, 630090 Russia*

[6]*Institute of Automation and Electrometry, Siberian Branch of the Russian Academy of Sciences, Novosibirsk, 630090, Russia*

[*]Corresponding author: yjrao@uestc.edu.cn

**Pulse generation often requires a stabilized cavity and its corresponding mode structure for initial phase-locking. Contrastingly, modeless cavity-free random lasers provide new possibilities for high quantum efficiency lasing that could potentially be widely tunable spectrally and temporally. Pulse generation in random lasers, however, has remained elusive since the discovery of modeless gain lasing. Here we report coherent pulse generation with modeless random lasers based on the unique polarization selectivity and broadband saturable absorption of monolayer graphene. Simultaneous temporal compression of cavity-free pulses are observed**

**with such a polarization modulation, along with a broadly-tunable pulsewidth across two orders of magnitude down to 900 ps, a broadly-tunable repetition rate across three orders of magnitude up to 3 MHz, and a singly-polarized pulse train at 41 dB extinction ratio, about an order of magnitude larger than conventional pulsed fiber lasers. Moreover, our graphene-based pulse formation also demonstrates robust pulse-to-pulse stability and wide-wavelength operation due to the cavity-less feature. Such a graphene-based architecture not only provides a tunable pulsed random laser for fiber-optic sensing, speckle-free imaging, and laser-material processing, but also a new way for the non-random CW fiber lasers to generate widely tunable and singly-polarized pulses.**

Random lasers have received considerable attention due to the unique properties that distinguish them from conventional resonant cavity lasers, including modeless, cavity-less, low coherence output, simple design, and good reliability operation [1-3]. Recently random fiber lasers based on the Rayleigh scattering and Raman effects in standard optical fibers have been advanced [4-6]. These fiber lasers with high quantum efficiency and large spectral tunability open up a new direction in both science and fiber-optic technologies. Reliable high-power operation, long-distance distributed fiber-optic sensing, and speckle-free imaging are promising novel areas for random fiber laser applications [7-10]. The development of pulsed random fiber lasers can also open up new applications, including ultrafast optical communications, precision sensing, high-resolution imaging, and next-generation laser processing and medical applications [11-13]. However, since random fiber lasers are modeless, it has been a fundamental scientific challenge to achieve mode-locking or to generate short pulses in such systems. Conventional methods of pulse generation in fiber resonators, such as mode-locking, spatial interference, and $Q$-switching, are generally not applicable to random fiber lasers with random feedbacks [14-15].

Graphene is an attractive two-dimensional material with a variety of exceptional electronic and photonic properties, and has been proposed in various applications [16-18]. Due to its unique band structure, graphene serves as an excellent saturable absorber [19-21]. Accordingly, effective graphene-based mode-locked lasers and all-optical fast modulators have been proposed [22-25]. As a waveguide, graphene also provides exceptional properties in polarization dependent transmissions [26-29], serving as a natural polarizer [30-32]. Moreover, compared to other materials for optical applications, such as carbon nanotubes or cadmium compounds [33-34], graphene has some merits: (1) broadband high nonlinearity with ultrafast response [35-36]; (2) high-power sustainability [14, 20-22], (3) high mechanical strength and durability [37], and (4) compatibility with silicon waveguide and fiber hybrids [36, 38-39]. With graphene-based hybrids as a static absorber, mode-locking and *Q*-switching have been examined [31-32], albeit still requiring a conventional resonant cavity lasing structure.

Here we report graphene-based pulse generation with cavity-free random fiber lasers, for the first time. By dynamically modulating the polarization via a polarization rotator (PR) and then transmitting through a designed graphene-coated *D*-shaped fiber (GDF), the continuous wave (CW) from the arbitrarily-polarized random fiber laser is transformed into coherent sub-nanosecond pulses even in the non-resonant single-pass configuration. The PR and the GDF hybrid waveguide serves simultaneously as the linearly polarized pulse generator (polarization modulation) and the pulse reshaper (saturable absorption). The pulses generated from the GDF not only inherit merits of the CW random fiber laser such as a high pump-power Stokes conversion efficiency, but also offer unique advantages such as a power-dependent singly-polarized output with polarization extinction ratios up to 41dB, and wide tunability in *both* pulsewidths and repetition rates, across two and three orders of magnitude respectively.

**Results**

**Architecture of the pulse generator.** Figure 1(a) illustrates the concept of generating highly polarized pulses for a CW random fiber laser. Determined by the graphene's anisotropy, the loss on the TM-polarization is much higher than that on the TE-polarization [28, 38]. When light transmits along the GDF, as shown schematically in the inset, it can be regarded as a *x*-polarization transmission filter. With the *x*-polarization periodically modulated by the PR, the transmission is a sinusoidal-like temporal wave after passing the GDF. Simultaneously, with Pauli blocking and self-phase modulation at high input powers [21-22, 40], the temporal width of sinusoidal-like wave can be dramatically compressed to form narrow pulses. In this process, the polarization rotation rate determines the pulse repetition rate while the saturable absorption efficiency determines the pulsewidth. Theoretical analysis is detailed in Supplementary Information Section S1.

With the finite element method (FEM), an example modeled $E_x$-field distribution is shown in Figure 1(b) with other polarization and fields shown in Supplementary Information Section S1. For the *D*-shaped fiber without graphene coverage, the *x*-polarized component and the *y*-polarized component have almost equal intensity. In comparison, with graphene cladding, the electromagnetic field distribution is pulled close to the graphene film and the *x*-polarized component is much stronger than the *y*-polarized component. With the graphene-based evanescent field enhancement, the transmitted CW random lasing field interacts slightly with the graphene film (detailed in Supplementary Information Section S1).

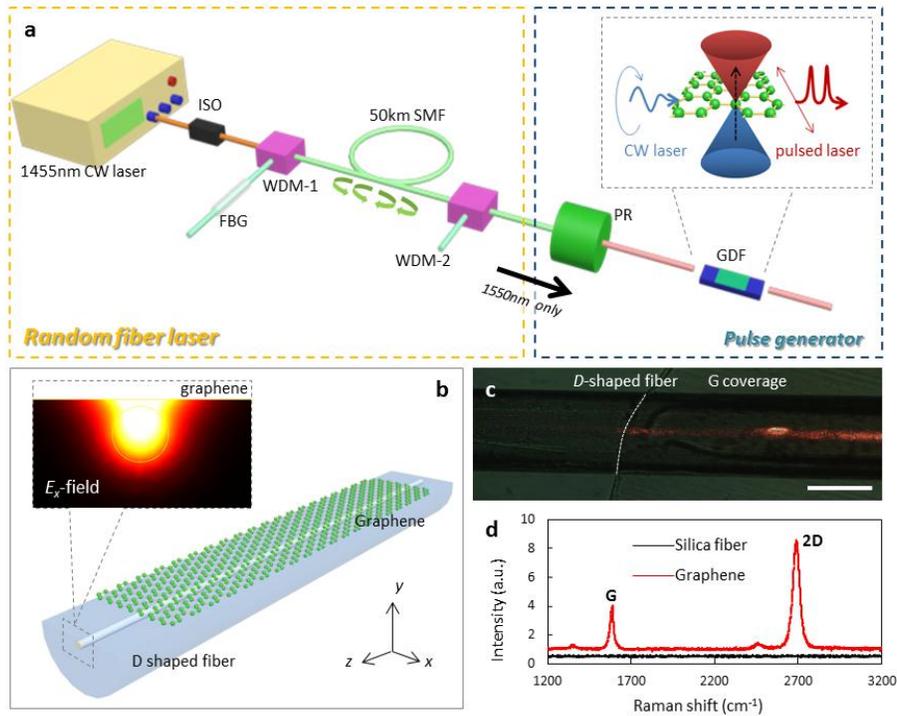

**Figure 1 | Widely tunable and singly-polarized pulse generation for random fiber lasers. a,** Schematic of the short pulse generation. A CW random fiber laser (yellow box) is polarization-rotated by a polarization rotator (PR), followed by intensity modulation

with the graphene-coated *D*-shaped fiber (GDF). **b,** Structure of the graphene-coated *D*-shaped fiber, with a graphene monolayer (green atoms) deposited on the top-side polished surface of a *D*-shaped fiber. The coverage length of the graphene $L_G$ is 16.7 mm and the polished depth of the *D*-shaped fiber *d* is 57 μm. Inset: Simulated $E_x$-field distribution, with the graphene film over cladding. **c,** Optical micrograph of the graphene-coated *D*-shaped fiber. The boundary of the graphene coverage is illustrated with the white dashed line. By coupling 633nm light into the fiber, the graphene-enhanced scattering is illustrated in the red scattering regions. Scale bar: 50 μm. **d,** Raman spectra of the CVD-grown graphene monolayer and the *D*-shaped silica fiber. The weak *D*-peak, the FWHM of the *G*-peak (32.5cm$^{-1}$) and *2D*-peak (38.2cm$^{-1}$), and the *G*-to-*2D* peak ratio (0.36) means that the graphene film is of monolayer and uniformity.

**Graphene-coated *D*-shaped fiber.** The *D*-shaped fiber was carefully polished with good uniformity and small surface roughness, with an insertion loss less than 10 dB over a polished length of 20 mm. Detailed fabrication process is shown in Supplementary Information Section S2. The fabrication of the GDF with a long length and a highly uniform surface is crucial to ensure effective light-graphene interaction and low scattering losses. Significantly, the polarization dependent loss (PDL) determines the initial polarization extinction ratio (PER) and the peak-to-noise ratio (PNR) of the generated pulses, while the absorption efficiency dominates how strongly the pulses could be modulated optically. Figure 1(c) shows an optical micrograph of the graphene cladded *D*-shaped fiber, under 633 nm transmission for clarity. In order to prepare high-quality graphene film and transfer it to the *D*-shaped fiber properly, monolayer CVD graphene was used [41] instead of exfoliated graphene. With the wet-transfer technique [36], we successfully covered graphene over several centimeters of the *D*-shaped fiber surface with good flatness and uniformity, even when the fiber was polished down with 58 μm depth (details in Supplementary Information Section S3). Fig. 1(d) shows the Raman spectrum of the GDF, which verifies that the deposited graphene is of sufficient quality for our measurement. For a GDF with 1.67 cm length, the excess graphene loss is measured to be ~ 22 dB.

**Random fiber lasing.** A high-power CW laser at 1455nm serves as the pump for the random fiber laser. After the accumulation of Rayleigh scatterings and Raman effects in a 50km single mode fiber (SMF), a CW random fiber laser at 1550nm is obtained. Random fiber lasers with other wavelengths are also achievable from their stimulated Raman scattering nature [42]. A fiber Bragg grating (FBG) serves as a mirror to reduce the lasing threshold [43], and a WDM is used to filter off the pump component (detailed in Supplementary Information Section S4). Fig. 2(a) shows the dependence of the laser power on the pump power. The lasing threshold is 0.88 W. When the pump power reaches 2.5 W, the output power of the random fiber laser measured after the WDM-2 is 160 mW. Considering the loss of the 50 km long system is over 10 dB, the pump-to-Stokes conversion efficiency of the random fiber laser is estimated conservatively to be above 60%, which could be even higher with short fiber lengths [42].

**Pulse generation and characteristics.** The polarization output of the random fiber laser is next modulated periodically using the PR, with a rotation speed tunable from 1 kHz to 3 MHz, prior to launching into the GDF. The power-dependent polarization selectivity and saturable absorption of the GDF are shown in Fig. 2(b). With a broadband tunable laser with a low average power of 9.8mW, the GDF has demonstrated its natural PDL as shown in the inset of Fig. 2(b). In the window of 1510 nm to 1570 nm, the transmission in the *x*-polarization was approximately -22dBm, while the transmission in the *y*-polarization was approximately -57dBm for a PER in excess of 34 dB. When the launched power of the GDF increases gradually, saturable absorption begins to occur along the GDF. Supplementary Information Section S5 details the saturable absorption and theoretical modeling. When the launched power of the *x*-polarized laser launched in the GDF increased from 1mW to 2.9W, the transmittance of the GDF increases from ~0.05% to ~ 3.4%. Contrastingly, for the *y*-polarized laser with launched power of 2.9W, the transmission stays predominantly constant with only a 0.0002% increase, within the noise level and negligible. This indicates that much higher power is needed to use the *y*-polarized light to saturate the GDF. Such a polarization-dependent saturable absorption is determined compositely by the polarization dependent transmission of the waveguide, and the polarization dependent nonlinear response of graphene.

The polarization properties of the graphene based random fiber laser are shown in Fig. 2(c), illustrating the measured Stokes vectors before the PR, after the PR, and after the GDF. Details of the polarization selectivity are shown in Supplementary Information Section S6. Initially, the CW random fiber laser has a low polarization degree of less than 30%, with a random polarization state. The PR then polarizes it and rotates its polarization, with subsequent polarized transmission filtering by the GDF.

The intensity dynamics after the PR are measured by a real-time oscilloscope, comparing before and after the GDF. Under 2.8 MHz rotation, Fig. 2(d) shows the transmission with increasing powers up to 2.88 W where a constant CW transmission is observed before the GDF. In contrast, after the GDF, as shown in Fig. 2(e), the CW laser becomes a sinusoidal-like oscillation. With increasing launched powers from 13.8 mW to 2.88 W, the sinusoidal-like oscillation is compressed to be sub-ns pulses gradually, at a constant repetition rate.

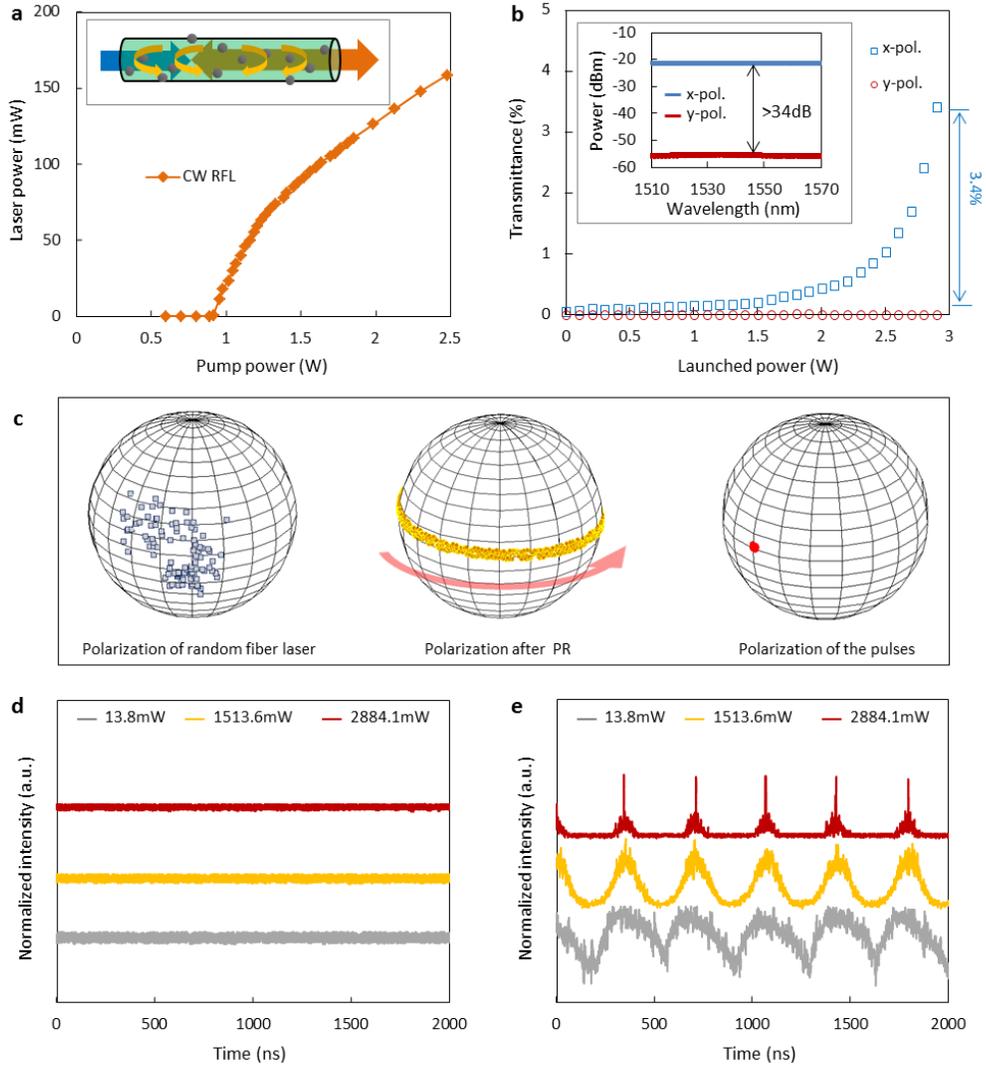

**Figure 2 | Polarization selection and pulse generation of random fiber lasers. a,** Output power of the CW random fiber laser. Inset: Distributed feedback of random fiber laser. **b,** Polarization dependent transmittance of the graphene-coated *D*-shaped fiber at 1550 nm under CW driving (blue squares: *x*-polarization; red circles: *y*-polarization). Inset: transmission spectra under 9.8 mW (blue curve: *x*-polarization; red curve: *y*-polarization). **c,** Poincaré map of the CW random fiber laser before the polarization rotator *PR* (low polarization degree), after the *PR* (rotated polarization), and after the graphene-coated *D*-shaped fiber (linear polarization). **d** and **e,** Intensity dynamics of the random fiber laser measured before and after the graphene-coated *D*-shaped fiber at 13.8 mW (grey curve), 1.51 W (yellow curve), and 2.88 W (red curve), respectively. The pulse repetition period in (**e**) is ~ 360 ns.

Fig. 3(a) shows the progression of the generated pulses under increasing launched powers and a fixed 2.8 MHz repetition rate. The (left column) blue data points are the measured pulsewidths while the (right column) yellow curves are from the theoretical calculations, based on the theoretical analysis in Supplementary Information Section S1. With the launched power increasing from 13.8 mW to 2.88 W, the sinusoidal-like waveform with full width at half maximum (FWHM) more than 180 ns is compressed to be narrow pulses gradually. The pulse generation becomes obvious when the launched power higher than 2.2 W. At the launched power of 2.88 W, the duty cycle is lower than 0.3%, where the duty cycle means the value of the FWHM of a single pulse comparing the temporal period of the PR. Fig. 3(b) demonstrates the ~ 900 ps pulse in detail. With the high power induced saturable absorption, the pulses are generated by compressing the sinusoidal-like waveform more than two orders of magnitude.

The correspondingly pulse spectra is shown in Fig. 3(c). At higher pulse energies based on the higher launched power, chirp and self-phase modulation is evident. At 13.4 mW initially, the 3-dB full-width half-maximum of the random fiber laser spectrum is ~ 0.1 nm, since the FBG performs as a narrow filter. At 2.88 W, the 3-dB spectral width increases to 0.96 nm and is dominated by the graphene pulse reshaping. Supplementary Information Section S7 also notes the 30-dB spectral width to illustrate the self-phase-modulation and chirp effects. Here the pulse compression and spectra broadening is modeled via the beam analysis and Fourier transform methods. The GDF formed pulses and details of the numerical modeling are illustrated in Supplementary Information Section S1.

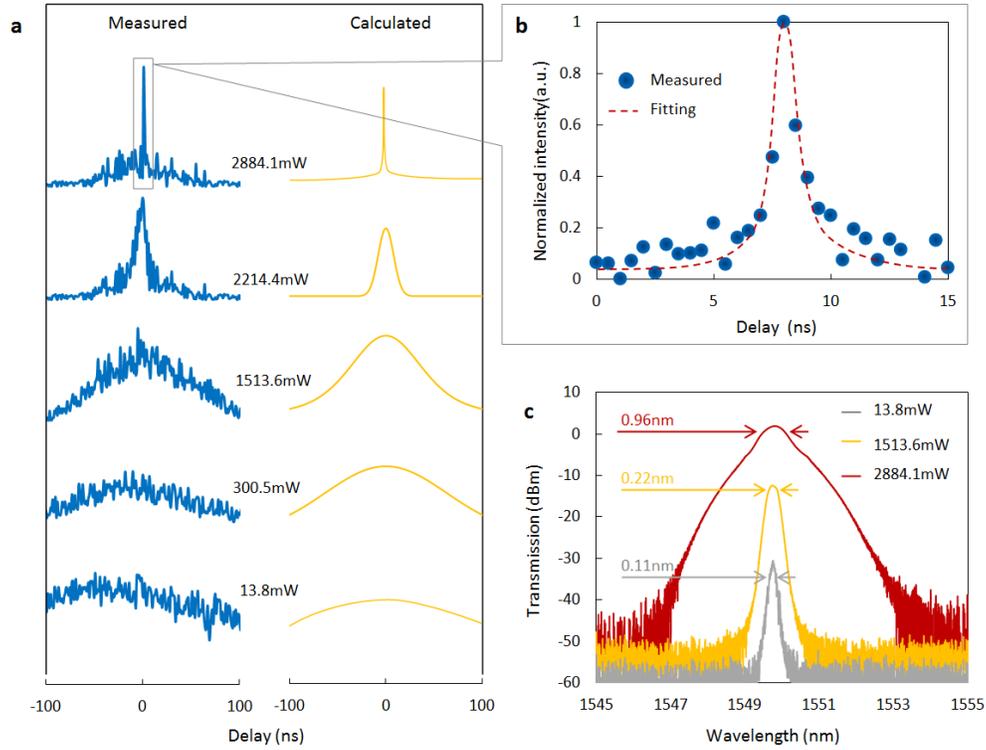

**Figure 3 | Temporal-spectra structure of the graphene-based pulse generation. a,** Measured (left column) and calculated (right column) temporal profiles of the generated pulses at 1549.5 nm and with a 2.8 MHz repetition rate. Different launched powers are examined to illustrate the pulse formation. **b,** Zoom-in of the pulse at 2.88 W. Raw measurements are illustrated in the blue data points, with a Gaussian fit in the black dashed line. The measured pulsewidth is estimated to be ~ 900 ps. **c,** Measured spectra of the pulsed graphene random fiber laser at launched power of 13.8 mW (grey), 1513.6 mW (yellow) and 2884.1 mW (red). The spectral 3-dB width increases from 0.11 nm to 0.96 nm, and correlates with the shortened temporal pulsewidths.

**Tunability.** Fig. 4(a) shows the dependence of the compressed-width and spectra for increasing the launched power of the GDF and at a fixed repetition rate, tuning over two orders of magnitude in the pulsewidth up to 2.88 W, with a fixed repetition of 2.8 MHz. With launched powers in excess of 2.88 W, we note that thermal damage of graphene starts to appear, limiting further compression currently. Moreover, considering the polarization-dependent saturable absorption of the GDF, the PER of the pulses is also power-controllable, as shown in inset of Fig. 1(a). With the launched power increasing from 1 mW to 2.88 W, the PER of the pulses increases gradually, due to stronger saturation in *x*-polarization. The PER reaches ~41 dB at ~2.88 W, one of the highest values observed for high power polarizers.

Fig. 4(b) shows the tuning of the repetition rate through the PR rotation rate, for a fixed launched power at 2.88 W, illustrated at 160 kHz, 560 kHz, and 3 MHz, for example. By increasing repetition rate, the sinusoidal-like pedestal is suppressed dramatically, and the pulse quality increases to 50.9%, at the launched power of 2.88 W and the repetition rate of 3 MHz. The time jitter of the pulse train generated by the PR and the GDF is mainly determined by the stability of the PR, whose temporal instability is lower than 0.1%. Hence, a higher repetition rate results in a lower time jitter.

Fig. 4(c) shows the pulsewidths versus the repetition rates, starting from 1 kHz to 3 MHz and with the pulsewidth depending exponentially on the repetition rate. A higher repetition rate also brings about narrower pulsewidth, as the pulses are compressed from the initial sinusoidal-like envelope formed by the GDF, based on its polarization selectivity. When the repetition rate increases, the width of the initial sinusoidal-like peaks becomes narrower. Hence, after compression, these faster sinusoidal-like peaks are transformed into even narrower pulses. Based on the numerical scaling, we note that the pulsewidth could reach sub-picosecond levels at a repetition rate above 6 MHz.

In addition, as the pulses are gradually compressed from the initial sinusoidal-like wave, a weak sinusoidal-like envelope is always persistent at the pulse pedestal, distinct from conventional cavity solitons or cavity mode-locking. Such a residual pedestal would deteriorate the pulse quality, which could be described as $Q_P = E_{FWHM}/E_0$, here $E_{FWHM}$ is the energy in the full width at half maximum (FWHM) of the pulse, $E_0$ is the energy of the pulse integrated over its period [44]. However, by increasing launched power or repetition rate, the sinusoidal-like pedestal could be dramatically suppressed, so that the pulse quality could be improved, as shown in Fig. 4(d). Referring Supplementary S1, calculated curve suggests that the compressed pulses can have a $Q_P$ >70%. However, limited by the background noise and instability of the PR, the measured pulse quality is relatively lower. In the experiment, the maximum measured $Q_P$ is 50.9%, with launched power 2.88 W and repetition rate 3 MHz.

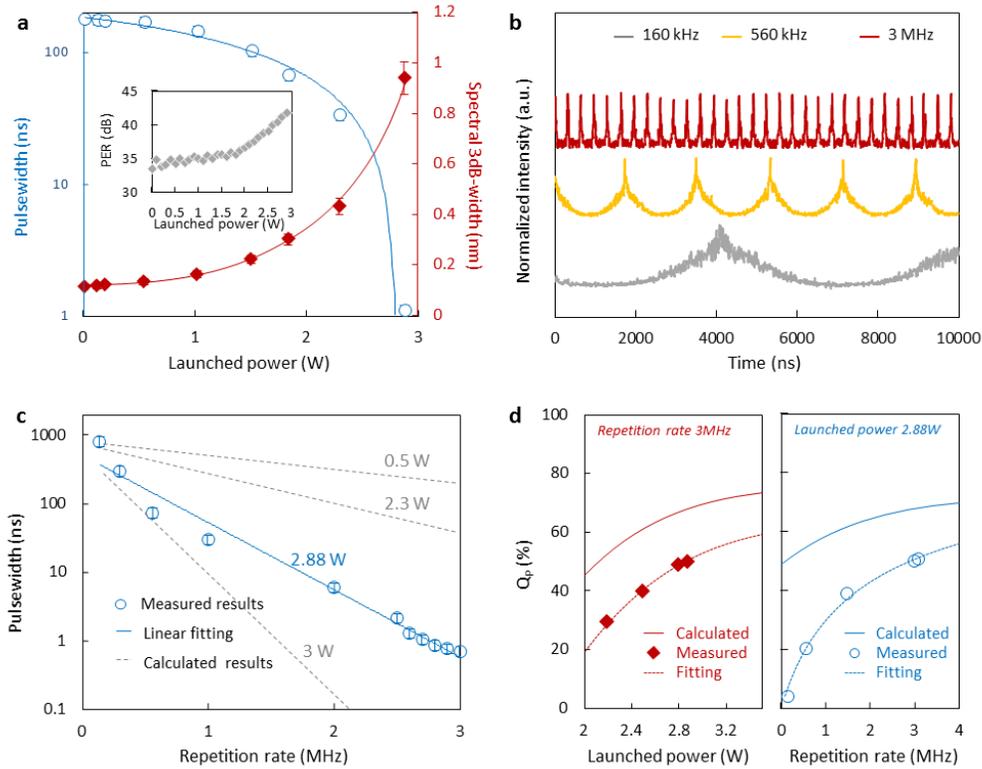

**Figure 4 | Characteristics and tunability of the generated pulses. a,** Pulsewidths (blue circles) and spectra width (red diamonds) for increasing launched power at 1549.5 nm. Inset: Power-dependent polarization extinction ratio at 1549.5 nm. **b,** Intensity dynamics at 160 kHz, 560 kHz, and 3 MHz repetition rates at 2.88 W and 1549.5 nm. **c,** Pulsewidth dependence on repetition rate, with blue circles representing the measurement data including error bars. The grey dashed lines are the theoretical models for different launched powers. **d,** Pulse quality dependence on launched power (Left side, with a fixed repetition rate 3 MHz), and pulse quality dependence on repetition rate (right side, with a fixed launched power 2.88 W).

## Discussion

Distinct from conventional mode-locking lasers, pulse generated by the single-pass GDF configuration is compressed from sinusoidal-like envelope. To achieve short pulsewidth and acceptable pulse quality, relatively high launched power and fast PR is necessary. However, to generate pulses with cavity-free random fiber lasers, external methods are necessary. Fortunately, the GDF based pulse generation has exceptional properties, and could be further improved in the future, by optimizing the smoothness/uniformity of the GDF, the performance of the PR, and increasing the launched power or the modulation speed of polarization.

In addition, beyond the random fiber lasers, the graphene based pulse generator could also be widely applied on typical non-random CW lasers in common fiber systems, as the polarization selectivity and the saturable absorption of the GDF is unique and universal. Hence, it provides a new way for the non-random CW lasers to generate widely tunable and singly-polarized pulses.

In summary, we have demonstrated a graphene based 900ps and widely-tunable pulse generation, with up to 41 dB polarization extinction for random fiber lasers. The single-pass configuration affords unprecedented widely-tunable pulses over two orders of magnitude in pulsewidth and three orders of magnitude in repetition rate, drawing from the synergistic advantages of graphene photonics and random fiber lasers. This not only inspires exploration of the full potential of graphene-based fiber devices for ultrafast photonics, but also the realization of pulsed random fiber lasers towards sensing, imaging and laser-material processing applications.

## Methods

**Theoretical analysis and numerical simulations.** The electric field distribution of the graphene-coated *D*-shaped fiber (GDF), the polarization-dependent transmission, and fast pulse generation were theoretically investigated and numerically simulated. To estimate the power density in the GDF, the field distribution of the graphene-coated *D*-shaped fiber was simulated by *COMSOL* (considering the index of graphene to be determined by its conductivity) by applying $\varepsilon_{g,eq}=-\sigma_{g,i}/\omega\triangle + i\sigma_{g,r}/\omega\triangle$ and $\omega^2\mu_0\varepsilon=n^2k_0^2$; see Supplementary S1.1. Moreover, modulated by the graphene-coated *D*-shaped fiber, the CW light from the random fiber laser would be transformed into highly polarized pulses with power that varies with time.

**GDF fabrication and characterization.** The GDF samples were fabricated by following steps. Firstly, graphene was grown by using the CVD method on Cu foil. Secondly, after removing Cu by $FeCl_3$ solution, a soft and flexible PMMA/graphene was prepared. Thirdly, the PMMA/graphene hybrid was covered on the polished surface of a *D*-shaped fiber, which was carefully fabricated with the polished depth of 58μm and polished length ~20 mm. Finally, the PMMA was removed by acetone vapor. The *D*-shaped fiber and graphene were clearly identified and characterized using optical microscopes (OPM), scanning electronic microscopes (SEM), and a Raman spectrum analyzer. More details are shown in Fig. S2 and Fig. S3.

**Measurement and simulation of the polarization dependent saturable absorption.** A polarization controller (PC) is used to adjust and fix the input polarization. The power launched into the graphene-coated *D*-shaped fiber can reach 3.5 W. The power meter with resolution of 0.1 dBm was used to detect the output power at 1550 nm. The transmittance could be calculated as $P_B/P_A$. Fig. S5 in Supplementary Information shows the method and the results.

**GDF based polarization selectivity measurement.** To verify the broadband polarization selectivity of the GDF, an experimental setup was built as shown in Supplementary Fig. S6. Light from 1510 nm to ~ 1570 nm was launched from a tunable fiber laser (81960A, Agilent, USA), and measured by a high resolution OSA (8163B, Agilent, USA) and a power meter. The output power of the laser is fixed at 9.8 mW. A polarizer was used to control the launched polarizations.

**Acknowledgements**

The authors are grateful for the equipment support from Prof. B. J. Wu, and helpful discussions from Prof. M. Sumetsky, and Dr. Shuwei Huang. This work is supported by National Natural Science Foundation of China under Grant 61290312, 61205048, 61107072, 61107073, 61475032, the Program for Changjiang Scholars and Innovative Research Team in Universities of China (IRT1218), the 111 Project (B14039).

**Author contributions**

B.C.Y. and Y.J.R. designed this work, B.C.Y. finished the numerical simulations. B.C.Y., H.W. and M.Q.F. performed the experiment, Z.N.W. and W.L.Z. provided the RFL. J.H.Z., Y.F.C. and Y.R.L. conducted the graphene deposition and transferring, X.L.C helped to prepare the D-shaped fiber samples, B.C.Y., Y.J.R., Z.N.W., Y.W., D.C., S.T., and C.W.W. discussed the work and prepared the manuscript.

**Additional information**

The authors declare no competing financial interests. Supplementary information accompanies this paper online. Reprints and permission information is available online at http://www.nature.com/reprints/. Correspondence and requests for materials should be addressed to Y.J.R.

# Supplementary Information

# Graphene based widely-tunable and singly-polarized pulse generation with random fiber lasers

B. C. Yao[1,2], Y. J. Rao[1*], Z. N. Wang[1], Y. Wu[1], J. H. Zhou[3], H. Wu[1], M. Q. Fan[1], X. L. Cao[1], W. L. Zhang[1], Y. F. Chen[3], Y. R. Li[3], D. Churkin[4,5,6], S. Turitsyn[4,5], and C. W. Wong[2]

*[1] Key Laboratory of Optical Fiber Sensing and Communications (Education Ministry of China), University of Electronic Science and Technology of China, Chengdu 610054, China*

*[2] Mesoscopic Optics and Quantum Electronics Laboratory, University of California, Los Angeles, CA 90095, United States*

*[3] State Key Laboratory of Electronic Thin Films and Integrated Devices, University of Electronic Science and Technology of China, Chengdu 610054, China*

*[4] Aston Institute of Photonic Technologies, Aston University, Birmingham, B47ET, United Kingdom*

*[5] Laboratory of Nonlinear Photonics, Novosibirsk State University, Novosibirsk, 630090 Russia*

*[6] Institute of Automation and Electrometry, Siberian Branch of the Russian Academy of Sciences, Novosibirsk, 630090, Russia*

[*] Corresponding author: yjrao@uestc.edu.cn

**S1. Theoretical analysis and numerical simulations.**

1.1 Electric field distributions of the graphene-coated *D*-shaped fiber.

According to the waveguide theory, the spatial distribution of mode field of the graphene-coated *D*-shaped fiber (GDF) is described in Eq. (S1) and Eq. (S2), which is determined by the effective refractive index of the GDF. Here $f_x$ and $f_y$ are the *x*- and *y*-polarized component of field intensity, $n_{eff}k_0$ is the propagation constant, $\omega$ is the photonic frequency, $\mu_0 = 4\pi \times 10^{-7}$ H/m, $\sigma_g$ is the conductivity of the graphene, $f_d(\varepsilon) = \{exp[(\varepsilon-\mu)/k_B T]+1\}^{-1}$ is the Fermi-Dirac distribution, $k_B$ is the Boltzmann's constant, *j* is the imaginary unit and *e* is the unit charge [S1-S3]. The distribution of the $n_{eff}$ of a graphene-coated *D*-shaped fiber is a function of both $\sigma_g$ and *D* (the polishing depth), which has been investigated in Ref. [S4-S5]. Accordingly, zoom-in images of the electric fields are shown in Fig. S1(a). The permittivity of the graphene is related to its conductivity $\sigma_g$ as $\varepsilon_{g,eq} = -\sigma_{g,i}/\omega\Delta + i\sigma_{g,r}/\omega\Delta$, where $\Delta = 0.4$ nm is the thickness of graphene. Hence, its index was calculated by applying $\omega^2\mu_0\varepsilon = n^2k_0^2$ [S5].

$$\begin{bmatrix} \dfrac{\partial^2 f_y}{\partial x \partial y} - \dfrac{\partial^2 f_x}{\partial y^2} \\ \dfrac{\partial^2 f_x}{\partial x \partial y} - \dfrac{\partial^2 f_y}{\partial x^2} \end{bmatrix} = (i\omega\mu_0\sigma_g + k_0^2 n_{eff}^2) \begin{bmatrix} f_x \\ f_y \end{bmatrix} \tag{S1}$$

$$\sigma_g = \frac{je^2(\omega - j/\tau)}{\pi\hbar^2} \left\{ \frac{1}{(\omega+j/\tau)^2} \int_0^\infty \varepsilon \left[ \frac{\partial f_d(\varepsilon)}{\partial \varepsilon} - \frac{\partial f_d(-\varepsilon)}{\partial \varepsilon} \right] d\varepsilon - \int_0^\infty \left[ \frac{f_d(-\varepsilon) - f_d(\varepsilon)}{(\omega+j/\tau)^2 - 4(\varepsilon/\hbar)^2} \right] d\varepsilon \right\} \tag{S2}$$

1.2 Graphene based highly polarized pulse generation and narrowing.

Considering the frequency of the polarization rotator (~MHz level) is orders slower than the frequency of the transmitting light (~200 THz), the average intensity of the CW random fiber laser in front of the GDF is regarded as a constant independent from time, the $E_I$. Rotated by the polarization rotator periodically, the components in $x$- and $y$-polarization before launched into the GDF are written as:

$$E_{x,0} = E_I \cos(\omega_P t) \tag{S3}$$

$$E_{y,0} = E_I \sin(\omega_P t) \tag{S4}$$

in which the $\omega_P$ is the rotating frequency. After propagating a distance of $z$ along the graphene-coated D-shaped fiber, the components in $x$- and $y$- polarization become:

$$E_{x,z} = E_I \cos(\omega_P t) \exp(\alpha_x ct/n_x) \tag{S5}$$

$$E_{y,z} = E_I \sin(\omega_P t) \exp(\alpha_y ct/n_y) \tag{S6}$$

wherein the $n_x$ and $n_y$ are the indexes of the GDF in $x$- and $y$- polarizations; the $\alpha_x$ and $\alpha_y$ are the attenuation coefficients in $x$- and $y$- polarizations, respectively. The $c$ is the light velocity in vacuum, and the $L_G$ is the length of the graphene-coated D-shaped fiber. Hence, the total transmission power at point $z$ is:

$$P_z = P_x + P_y = E_{x,z}^2 + E_{y,z}^2 = E_I^2 \left\{ \cos^2(\omega_P t) e^{2\alpha_x M_x} + \sin^2(\omega_P t) e^{2\alpha_y M_y} \right\} \tag{S7}$$

wherein the $M_x = ct/n_x$ and $M_y = ct/n_y$. Moreover, we take the polarization dependent saturable absorptions into consideration. The maximum power launched into the GDF fiber in our experiment is ~ 3 W. Referring the effective field distributions calculated in S1.1, the power density in the GDF in our experiment was much lower than 1 GW/cm$^2$, which is far lower from the saturated limitation for a graphene based waveguide [S6]. Accordingly, we approximately regard the polarization dependent saturable absorption based transmission as [S7]:

$$P_{SX} = A_x e^{B_x P_x} \tag{S8}$$

$$P_{SY} = A_y e^{B_y P_y} \tag{S9}$$

Here, $A_x(u_c,L_G)$, $A_y(u_c,L_G)$ and $B_x(u_c,L_G)$, $B_x(u_c,L_G)$ are constants determined by the length and Fermi-level of graphene, in which $u_c$ is the chemical potential of the graphene while $L_G$ is the length of the GDF.

In practice, the difference between the $M_x$ and the $M_y$ is negligible. Applying the *Matlab*, and adopting parameters $\omega_P = 2\times10^5\pi$ (100 kHz), $10^6\pi$ (500 kHz) and $2\times10^6\pi$ $\pi$ (1 MHz), the pulse series $P_z(t)$ is shown in Fig. S1 (b). These results are also corresponding to the Fig. 4 in the text. The values of the $\alpha_x$ and $\alpha_y$ are from the measured results in Ref. [S8]. Moreover, when fixing the repetition rate at 100 kHz, by adopting different $A_xA_y$ and $B_xB_y$, the pulsewidth is tunable, as shown in Fig. S1 (c). Fig. S1 indicates, by either increasing the saturable absorption ratio or the repetition rate, the pulses can be narrowed.

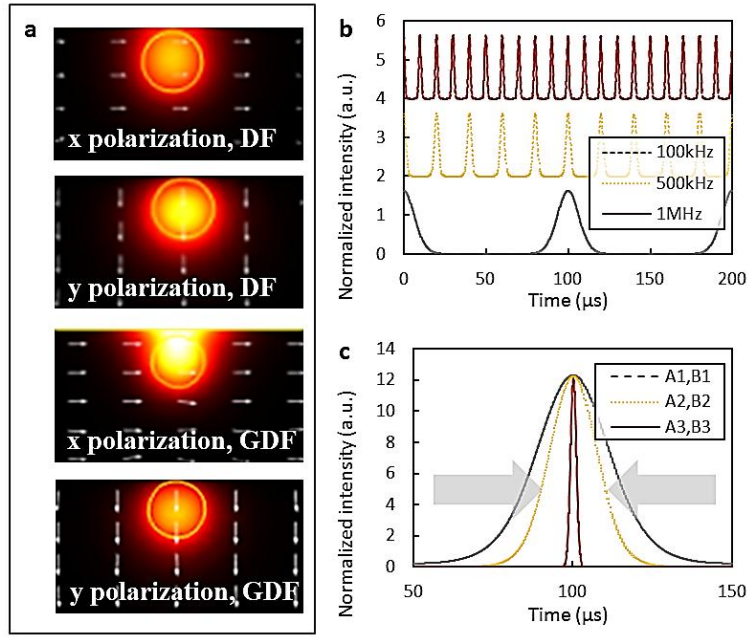

**Figure S1 | Numerical simulation results.** (a) Electric field distributions for the

*D*-shaped fiber (*DF*) and the graphene-coated *D*-shaped fiber (*GDF*). (b) Pulses generated by tuning the rotating speed at 100 kHz, 500 kHz, and 1MHz, with a normalized power. (c) Width of a single pulse, by adopting different parameters, the FWHMs could be 25 μs (grey), 12 μs (yellow), and 2 μs (red).

**S2. Fabrication of the GDF.**

GDF samples were fabricated by using the following steps shown in Fig. S2. Firstly, a soft and flexible PMMA/graphene was prepared; secondly, a specially designed *D*-shaped fiber was fabricated; thirdly, the PMMA/graphene hybrid was covered on the polished surface of the *D*-shaped fiber, finally, the PMMA film was removed. In the CVD process, graphene films were grown on Cu foils (Alfa Aesar, No. 13382), under 1000 °C, 50sccm $CH_4$ and 50sccm $H_2$ [S9]. A thin PMMA film was spin-coated on the surface of graphene/Cu foil, to form a PMMA/graphene/Cu sandwich like structure. The thickness of the PMMA was ~1μm. Then the underlying Cu foil was chemically etched by 1M $FeCl_3$ solution, the obtained PMMA/graphene film is of good flexibility. On the other hand, by using mechanical polishing method in wet environment, the side of a single mode fiber (SMF-28e, Corning Inc.) was polished into a *D*-shaped fiber. The polishing process was repeated carefully to reduce the scattering loss [S10]. Moreover, to ensure the interactions between the transmitting light and the graphene film, the polishing depth was accurately controlled to be ~58μm (±0.5μm), which means the distance between the core to the polished surface is less than ~0.5μm. Subsequently, the flexible PMMA/graphene film was washed in DI water 3 times and transferred onto the side surface of the *D*-shaped fiber, which had been ultrasonically cleaned in sequence by acetone, ethanol and DI water. Then the PMMA/graphene/D-shaped fiber was dried at room temperature for 12 hours and baked at 180 °C for 10 min. Finally, the PMMA was removed by acetone vapor.

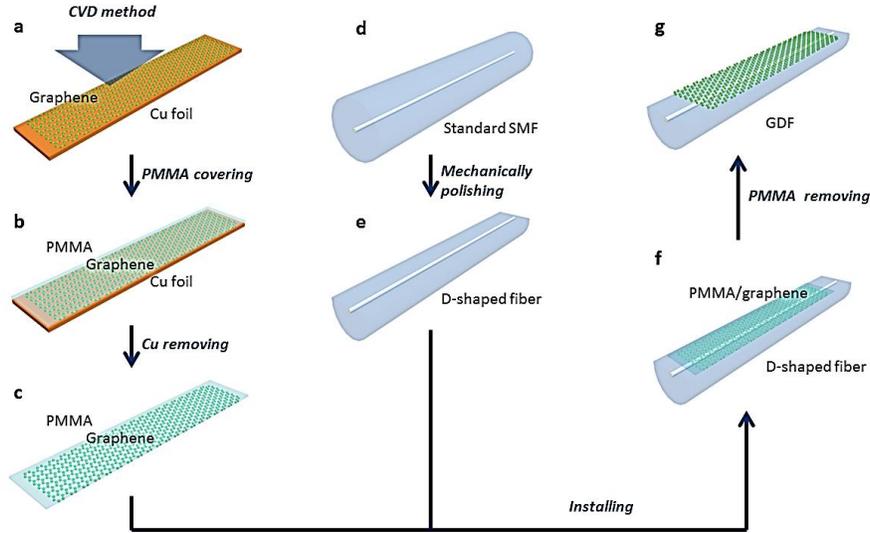

**Figure S2 | Fabrication process of the GDF. a,** By using the CVD method, a monolayer graphene was grown on a Cu foil. **b,** PMMA/graphene/Cu foil sandwich structure. **c,** PMMA/graphene flexible film. **d,** Standard SMF with core diameter 8μm and cladding diameter 125μm. **e**, By using mechanical polishing method, the standard SMF was fabricated to be a *D*-shaped fiber. **f**, Transfer of the PMMA/graphene onto the *D*-shaped fiber. **g,** PMMA removal.

**S3. Extended characterization of the GDF.**

Optical microscopes (OPM), scanning electronic microscopes (SEM) and Raman spectra were used to monitoring the fabrication process and the quality of the GDF sample, as shown in Fig. S3. Fig. S3(a) ~Fig. S3(e) show the OPM pictures for a typical as-fabricated GDF, with focusing on the boundary of the graphene coverage, with a series of amplification ratios. The OPM pictures demonstrate the combination of the GDF. Fig. S3(f) displays a SEM pictures of the sectional view and the uniform polished surface of the *D*-shaped fiber. The surface is smooth and uniform, and the polished depth is ~58 μm, i.e. ~0.5 μm away from the fiber core. Fig. S3(g) shows the Raman spectra of the graphene deposited on the *D*-shaped fiber, measured at 3 random positions, verifying that the graphene is of monolayer and high quality [S11-S13].

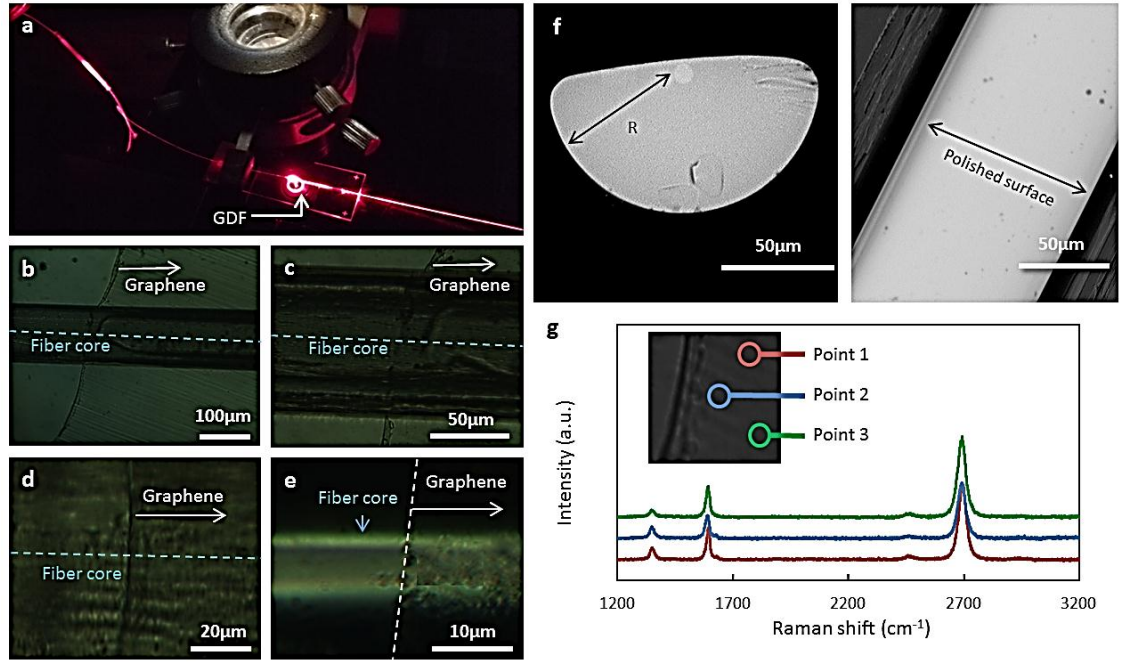

**Figure S3 | Characterizations the GDF. a,** A GDF with length of 16.7mm guiding 633nm light is illustrated under an optical microscope. **b** to **e**, Optical microscope images of the GDF. Here the bars in (**b**), (**c**), (**d**) and (**e**) are 100μm, 50μm, 20μm, and 10μm respectively. **f**, Scanning electron microscope image of the sectional view and the polished side of the *D*-shaped fiber. Here the scale bar is 50μm. **g,** SEM Raman spectra of graphene deposited on the *D*-shaped fiber measured from 3 different positions on the fiber, illustrating uniformity.

### S4. Measurement setup.

Devices and instruments arranged in the experimental system (briefly shown in Fig. 1 in text) are specifically described in Fig. S4. A CW laser at 1455 nm (KEOPSYS, France) was adopted as the pump, whose maximum output power could be tunable up to 10 W. The FBG written by a 248nm laser with reflectivity of 0.95 was used to decrease the pump threshold for lasing, its Bragg peak width is 0.5 nm. The polarization rotator (*PR*) includes: an attenuator (ATT), a polarizer, a polarization modulator (PM, General Photonics, USA), and an Erbium-doped fiber amplifier (EDFA). The speed of the PM is tunable from 1 kHz to 3 MHz. The launched power in front of the graphene-coated *D*-shaped fiber is tunable up to 3.5 W. Finally, the output light was measured by using an optical spectrum analyzer (OSA) (Ando, Japan), an oscilloscope (Tektronix, USA, maximum sampling rate 5 GHz) and a polarization analyzer (Agilent, USA)

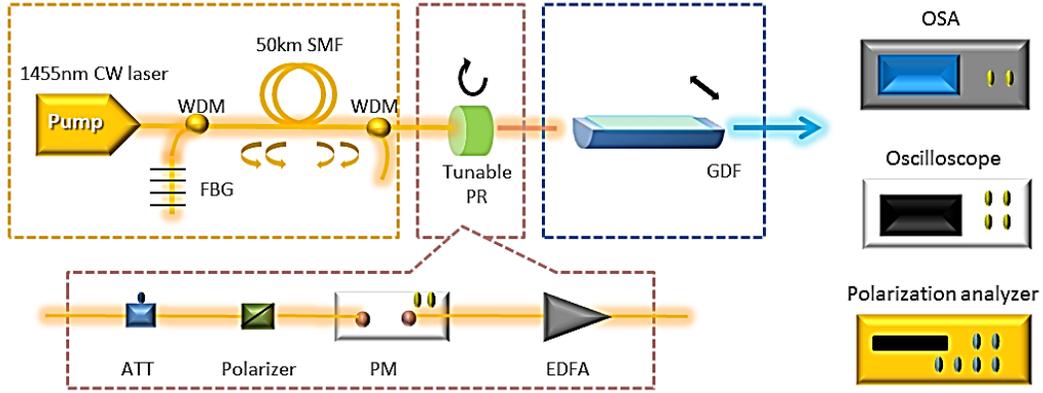

**Figure S4 | Measurement setup.** Here the CW random fiber laser, PR, and GDF are marked by yellow box, pink box, and blue box respectively.

**S5. Measurement and simulation of the polarization dependent saturable absorption.**

Saturable absorptions exist in the GDF. We measured the saturable absorption of the GDF, by using the experimental setup shown in Fig. S5(a). Here a polarization controller (PC) is used to adjust and fix the input polarization. The power launched into the graphene-coated *D*-shaped fiber can reach 3.5 W. The power meter with resolution of 0.1 dBm was used to detect the output power at 1550 nm. The transmittance could be calculated as $P_B/P_A$. Fig. S5(b) shows the transmissions of the *x*- and *y*- polarization measured by the power meter, *via* Channel *B*.

In addition, in this work, limited by the thermal damage threshold of the GDF, results under higher power (>3 W) are difficult to be provided. By using the formula $T = 1 - \Delta T \times exp(-P/P_{sat}) - T_{ns}$ [S14], we can approximately calculate the saturable absorption curve completely, as shown in Fig. S5(c). Here $\Delta T$ is the modulation depth, $P$ is the input power, $P_{sat}$ is the saturating power, and $T_{ns}$ is the non-saturable absorbance. Considering the Fermi-level tenability of graphene, it is also supposed that under an ultrahigh power, the transmission of *y*-polarization would be higher than the *x*-polarization [S15-S16].

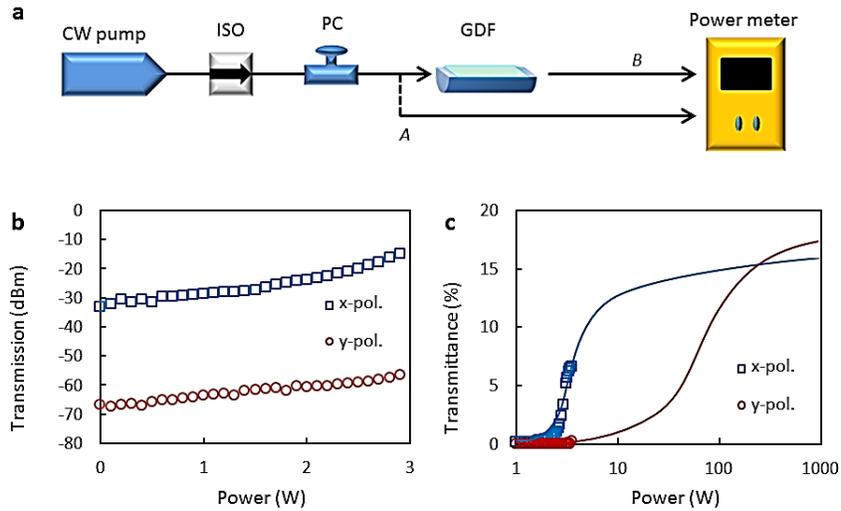

**Figure S5 | Saturable absorption measurement of the GDF. a,** Powers are measured before and after the GDF, so that the transmittance could be calculated precisely. **b,** Measured transmissions of the of the *x*- (blue boxes) and *y*- (red circles) polarization. **c,** Numerically calculated correlation of the power and transmittance of the GDF.

**S6. GDF based polarization selectivity measurement.**

According to its surface propagation properties, it has been verified that graphene could dramatically enhance the surface evanescent fields (see Fig. 1 in text) and induce a polarization dependent loss (see Fig. 2 in text) [S17-S20]. To verify the broadband polarization selectivity of the GDF, an experimental setup was built as shown in Fig. S6(a). Light from 1510 nm to ~ 1570 nm was launched from a tunable fiber laser (81960A, Agilent, USA), and measured by a high resolution OSA (8163B, Agilent, USA) and a power meter. The output power of the laser is fixed at 9.8 mW. A polarizer was used to control the launched polarizations. The measured spectra are shown in Fig. 2(a) in text.

  Moreover, Poincaré representation of the graphene pulsed random fiber laser measurements are shown in Fig. 6(b), before the PR (Point A), after the PR (Point B), and after the GDF (Point C). Modulated by the PR and the GDF, the CW random fiber lasing is tuned to highly-polarized pulses.

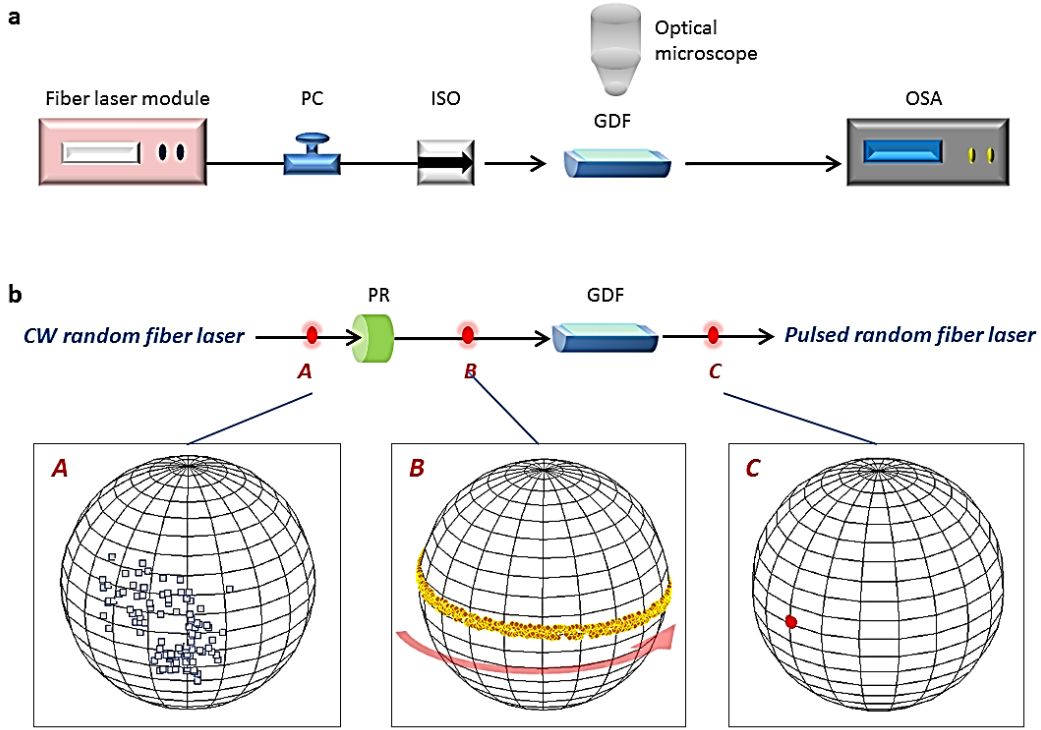

**Figure S6 | GDF works as a polarization selector. a,** Experimental setup to measure the polarization selectivity. **b,** Test points: before the PR (Point A), after the PR (Point B), and after the GDF (Point C).

**S7. Spectra of the pulsed random fiber laser.**

The nonlinear pulse reshaping of the pulses could also be verified by observing the spectral changes, considering the Fourier transformations [S21-S22]:

$$P_T(j\Omega) = \int_0^{+\infty} P_T(t) e^{-j\Omega} dt \tag{S10}$$

With increasing the power launched into the GDF, more complicated nonlinear effects appear, such as self-phase-modulations and chirps, which contribute the spectral broadening. Fig. S7(a) and S7(b) concludes the correlation of "Power vs 3dB-width" and "Power vs 30dB-width" for the graphene pulsed random fiber laser.

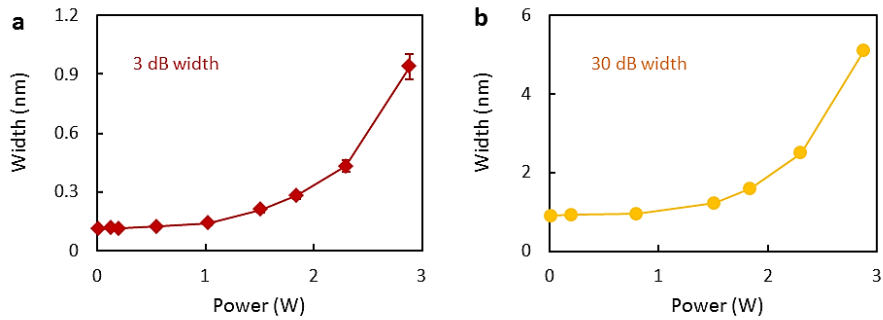

**Figure S7 | Spectral broadening of the graphene pulsed random fiber laser. a,** Power vs 3dB-width. **b,** Power vs 30dB-width.

**Supplementary References**